# H(T) PHASE DIAGRAM IN $Nb_3Sn$: A DIFFERENT BEHAVIOR IN SINGLE CRYSTALS, POLYCRISTALLINE SAMPLES AND MULTIFILAMENTARY WIRES

M. G. Adesso[1,2], D. Uglietti[1], R. Flükiger[1], M. Polichetti[2] and S. Pace[1]

[1]Institute of Applied Physics-GAP, University of Geneva,
Quai de l'Ecole de Médecine, 1211 Geneva, Switzerland

[2]Department of Physics, SUPERMAT, INFM, CNR, University of Salerno,
Via S. Allende, 84081 (Salerno) Italy

**ABSTRACT**

A comparative study of magnetic behavior in a DC magnetic field up to 19 Tesla has been undertaken on different $Nb_3Sn$ samples, i.e. single crystals, polycrystal samples and multifilamentary wires. From the 1st and 3rd harmonics of the AC magnetic susceptibility a peak effect was experimentally observed in the single crystal and in the polycrystal, whereas this phenomenon is absent in the measured wires. The corresponding H vs T phase diagram reveals a different behavior between wires on one hand and polycrystalline and single crystal samples on the other. In particular, from the 3rd harmonics it has been observed that vortex thermally activated flux creep phenomena are relevant in the wires, whereas the static critical state models are more appropriate to describe single crystals.

**KEYWORDS:** $Nb_3Sn$, Bragg Glass Phase, Peak Effect, harmonics of the AC magnetic susceptibility, H-T Phase Diagram

**PACS:** 74.70.Ad, 74.25.Ha, 74.25.Op, 74.25.Qt

## INTRODUCTION

$Nb_3Sn$ is one of the most attractive materials for the manufacture of high-field superconducting magnets. Since 1954, when superconductivity was observed for the first time [1], a strong interest has been devoted to this compound. $Nb_3Sn$ is a BCS low $T_c$ superconductor, crystallizing in the A15 structure [2], with a superconducting critical temperature $T_c$ = 18 K [1]. A structural transformation, occurring in a very narrow composition range only (at stoichiometry) [3], has been detected in pure samples (single crystals [4] and some polycrystals [3, 5]), from a high temperature cubic phase to a low temperature tetragonal phase [3]. Nevertheless, most of the studies on this compound are aimed at multifilamentary wires. In particular, the fabrication methods [6,7,8], the mechanical effects (effects of the stress on $T_c$ and critical current density, $J_c$ [9]) and the superconducting properties [10, 11] (enhancing of $J_c$ and $B_{c2}$ with addition of different elements) have been widely studied. In spite of this large scientific production, the field/temperature (H-T) phase diagram in $Nb_3Sn$ is not completely clear. The most important result on this subject has been obtained by Suenaga [12], who detected an irreversibility line, quite distinct from the $T_c$-line, in a commercial multifilamentary wire. Nevertheless, the H-T phase diagram in single crystals and polycrystals has not been analyzed in detail.

In this work, a comparative study on the magnetic properties of a single crystal, a polycrystal and a bronze route multifilamentary wire has been analyzed, by using both the 1st and the 3rd harmonics of the AC magnetic susceptibility. The corresponding H-T phase diagrams have been obtained both by a traditional way (from 1st harmonic measurements). In addition, we propose here a new method, based on the analysis of the 3rd harmonics.

**EXPERIMENTAL TECHNIQUE**

In the measurements of the harmonics of the AC magnetic susceptibility, a linear magnetic response corresponds to the detection of the 1st harmonics only, whereas non-linear magnetic properties are associated to the existence of higher harmonics [18, 19].

The measurement reported in this paper have been performed by a home-made susceptometer equipped with two different lock-in amplifiers devoted to the detection respectively of the 1st and 3rd harmonics as a function of the temperature ($T$) in the range [4.5-120 K], the DC magnetic field up to 19 Tesla, the frequency ($\nu$) from 10 to $10^3$ Hz, and the amplitude ($h_{AC}$) up to ~ 300 Oe of the AC magnetic field. In particular here we report harmonics measured at a fixed $\nu$ = 107 Hz and $h_{AC}$ = 128 Oe, at various DC fields, in the Zero Field Cooling configuration, acquired for increasing temperature. Both the AC and the

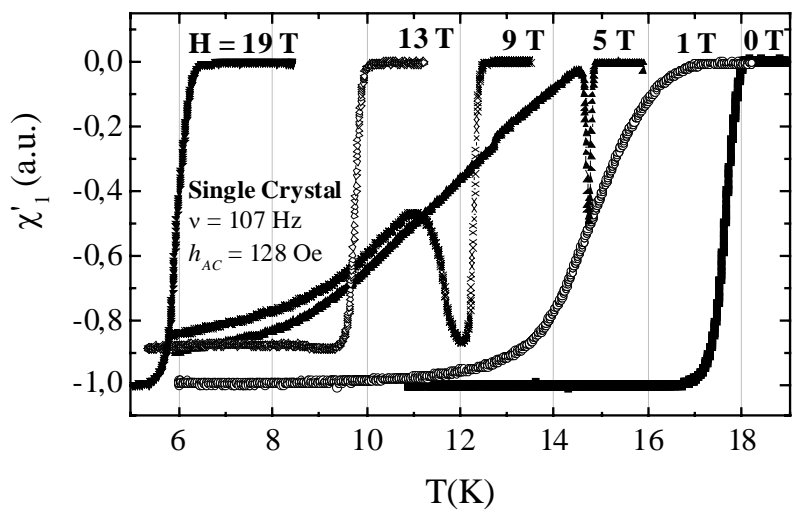

**FIGURE 1**. Real part of the 1st harmonics of the AC magnetic susceptibility, $\chi_1'(T)$, at various DC fields, measured on the single crystal sample.

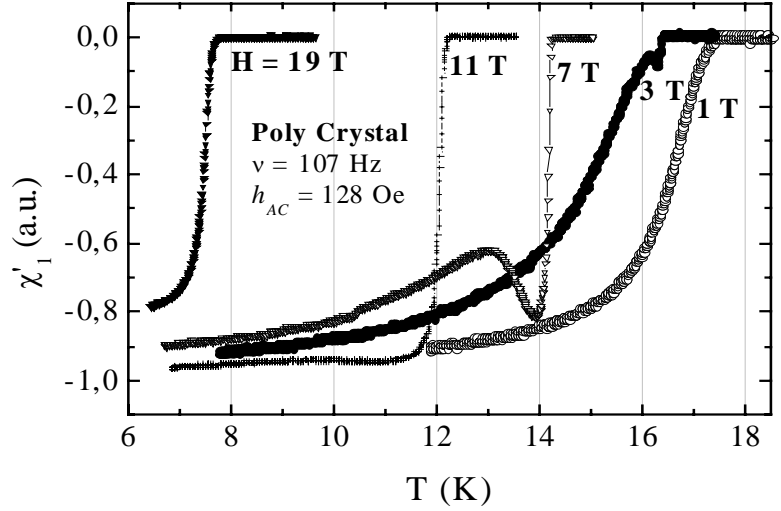

**FIGURE 2.** $\chi_1'(T)$ performed on the polycrystalline sample, at various DC fields. DC fields are applied parallel to the longer dimension of the sample.

## EXPERIMENTAL RESULTS

We analyzed different $Nb_3Sn$ samples, in particular: a single crystal, a polycrystal and a bronze route multifilamentary wire. More details about the analyzed samples are reported elsewhere [13], [14] and [8].

### 1st harmonics analysis

FIGURES 1, 2 and 3 show the real part of the 1st harmonics of the AC magnetic susceptibility as a function of the temperature, $\chi_1'(T)$, at various DC fields, for the single crystal, the polycrystalline sample and the multifilamentary wire, respectively.
A dip in $\chi_1'(T)$ has been detected in both the single crystal and in the polycrystalline sample for fields higher than 3 Tesla, up to a threshold field (10 Tesla for the polycrystal, 13 Tesla for the single crystal). This dip is directly associated to a peak in the critical current density, namely the "Peak Effect" [15, 16]. On the contrary, no dip has been detected in the measurements performed on the wire. In FIGURE 3, at T ~ 9 K in zero field it is also possible to observe an additional superconducting transition, associated with Nb in the wire [8], that it is not present any more at higher fields $H_{c2}(Nb)$ being under 1 Tesla [17].
The transition width is almost field independent in the single and in the polycrystalline samples for H > 3 Tesla, also at high fields, where the Peak Effect cannot be detected anymore. On the contrary, it becomes larger for increasing DC fields in the wire (as it generally occurs in the samples where no peak effect can be detected).

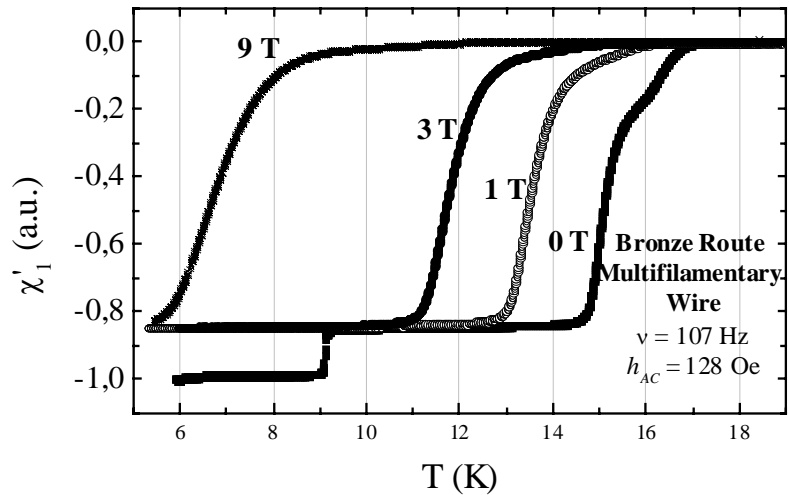

**FIGURE 3**. Temperature dependence of the real part of the 1$^{st}$ harmonics, for the Nb$_3$Sn wire, measured at various DC fields.

### 3$^{rd}$ harmonics analysis

We found that the common features between the polycristalline and the single crystal, opposite to the observed magnetic behavior in the wire, can be better evidenced by the analysis of the 3$^{rd}$ harmonics of the AC magnetic susceptibility.
The temperature dependence of the real part of the 3$^{rd}$ harmonics, $\chi_3'(T)$, at various DC fields, is illustrated in FIGURES 4, 5 and 6 for the three analyzed samples. The general features of the nonlinear magnetic response in the single and the polycrystalline samples are similar. In particular, $\chi_3'(T)$ is characterized in both the samples by a single positive peak near T$_c$, at fields lower than 3 Tesla, where no peak effect has been observed in the 1$^{st}$ harmonics. Two positive peaks can be identified at higher DC fields (corresponding to the peak effect detected by the 1st harmonic): the first one, near $T_c$, is almost field independent, whereas the other one tends to decrease for increasing fields and disappears at a certain DC field, where the peak effect cannot be detected any more in the 1st harmonics. The positive peak near $T_c$, instead, can be always observed, up to the maximum available field (19 Tesla).

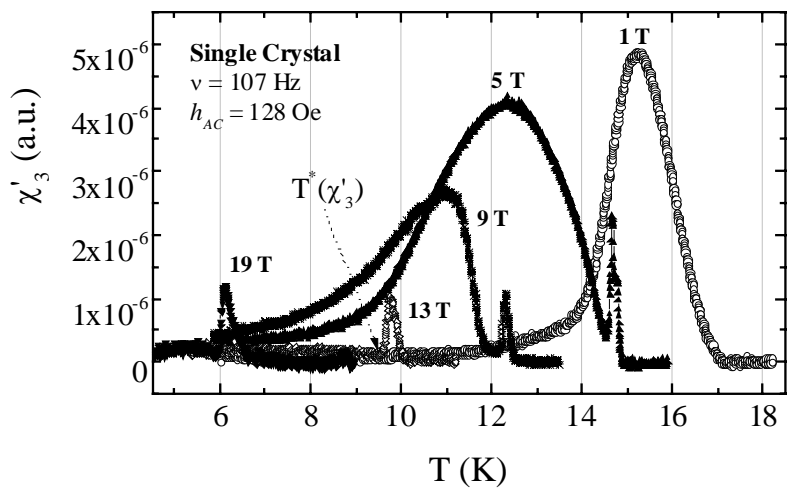

**FIGURE 4** $\chi_3'(T)$ measured on a single crystal at various DC fields.

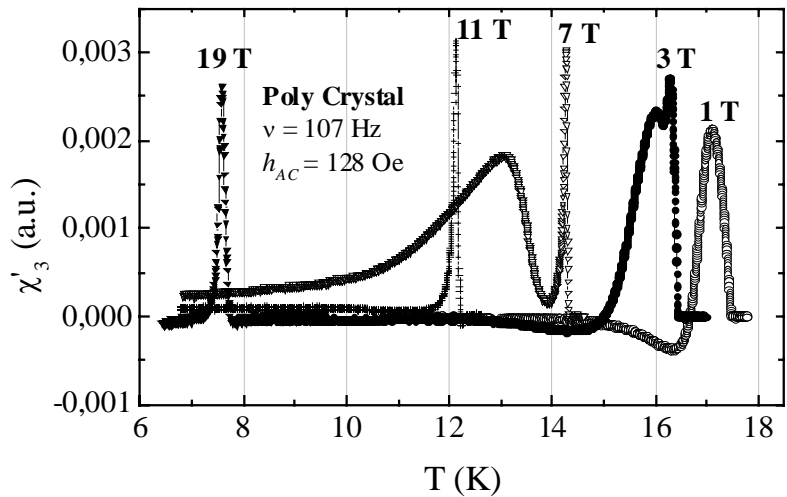

**FIGURE 5** $\chi'_3(T)$ measurements, performed on the polycrystalline sample at various DC fields.

In the wire sample, only one positive peak near $T_c$ has been detected in $\chi'_3(T)$ curves,

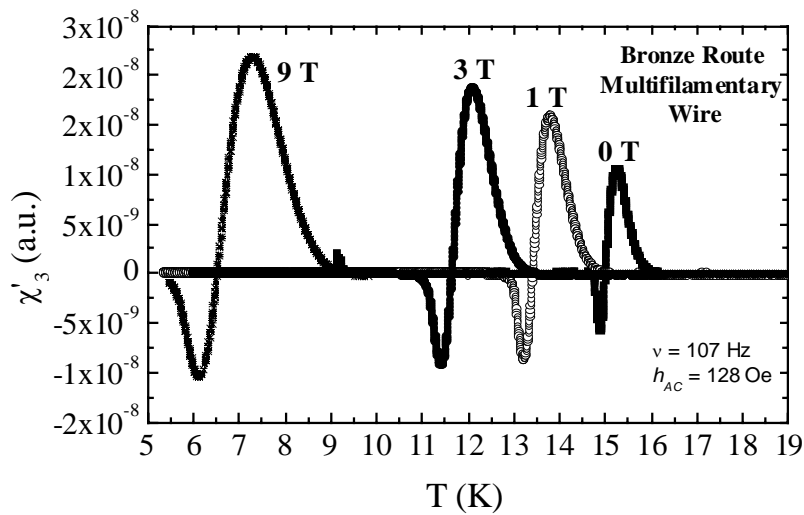

**FIGURE 6** Temperature dependence of the real part of the 3rd harmonics, measured on the wire sample, at various DC fields.

that is field dependent, e.g. its height grows for increasing field, contrarily to what observed in the polycrystal and in the single crystal.

Although some common features between the polycrystal and the single crystal have been evidenced, it is possible to observe some differences, too. In particular, we noticed that the relative heights of the peaks are different depending on the samples: in the single crystal, the height of the peak near $T_c$ is always lower than the second peak height, contrary to what happens in the bulk. The reason of this difference is not clear and theoretical/numerical investigations are needed. Moreover, we detected in the polycrystal sample a small negative bump at low fields (see H = 1 Tesla in FIGURE 5), in a temperature range far from $T_c$, absent in the single crystal. The peak is not in agreement with the static critical state models [18] predicting only non-negative values for $\chi_3'(T)$, but it can be explained in terms of the vortex dynamic phenomena [19] determined by thermally activated creep phenomena. Nevertheless, these dynamic phenomena tend to disappear for increasing fields. In contrast, the negative bump at low temperatures is always present in the multifilamentary wire (see FIGURE 6), its height growing in absolute value for increasing fields. This behavior is similar to that observed in the high $T_c$ superconductors [19] and can be associated to flux creep phenomena [20].

**H-T Phase Diagram**

In FIGURES 7 and 8 the H-T phase diagrams have been reported for both the single crystal and the polycrystalline sample, respectively. They have been obtained by plotting the superconducting transition temperature ($T_c$) and the temperature corresponding to the $\chi_1'(T)$ dip local minimum ($T_p$), determined by 1$^{st}$ harmonics measurements, at various DC fields. The presence of a Peak Effect indicates that a transition between a Disordered Phase and the Bragg Glass Phase is occurring, following T. Giamarchi [21]. In the region between the $T_c$-line and the $T_p$-line a disordered phase can be detected, whereas the H-T region for temperatures lower than $T_p$ is characterized by the Bragg Glass phase.

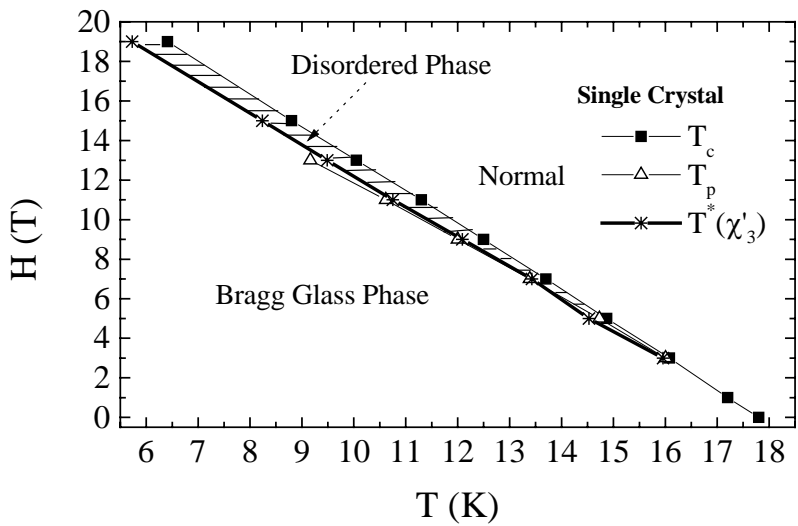

**FIGURE 7.** H-T phase diagram for the Nb$_3$Sn single crystal, obtained by both 1$^{st}$ and 3$^{rd}$ harmonics of the AC magnetic susceptibility.

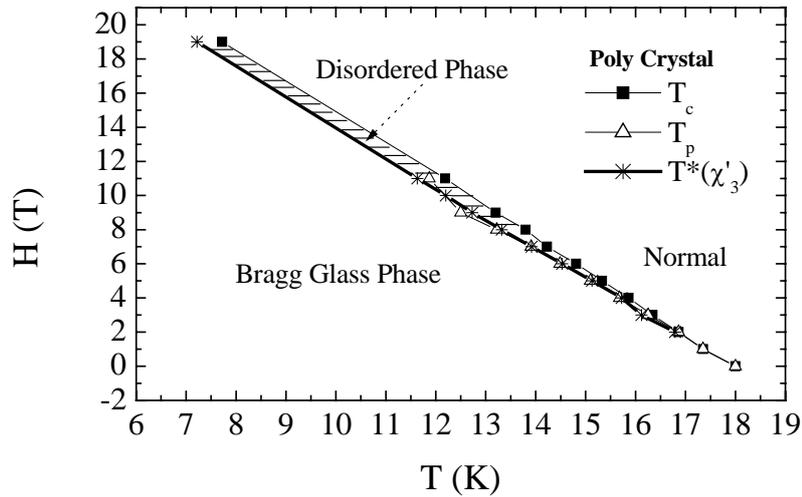

**FIGURE 8**. H-T phase diagram in the Nb$_3$Sn polycrystal.

In this work, it was found that the 1$^{st}$ harmonics is not appropriate to detect the transition between the Bragg and the disordered phase in the low temperatures/high fields range. For this reason, the $T^*(\chi_3')$ has also been plotted, corresponding to the onset temperature of the positive peak near T$_c$, in the real part of the 3$^{rd}$ harmonics. It has been already observed [22] that $T^*(\chi_3')$ can be detected up to the highest available field, giving up the possibilities to have an agreement with the theoretical models [21, 23] predicting the existence of the Disordered/Bragg phase transition at all the fields. The irreversibility line [24], obtained as the onset of the 3$^{rd}$ harmonics [25], corresponds to the T$_c$ line, so for simplicity it has not been plotted on the phase diagrams reported in Fig. 7 and 8.

In contrast, a clearly distinct irreversibility line can be identified in the H-T phase diagram of the wire, as shown in FIGURE 9. Nevertheless, no Bragg phase has been detected in the multifilamentary wire. The obtained phase diagram for the wire is in agreement to what reported in literature [12], but is different from both the phase diagrams

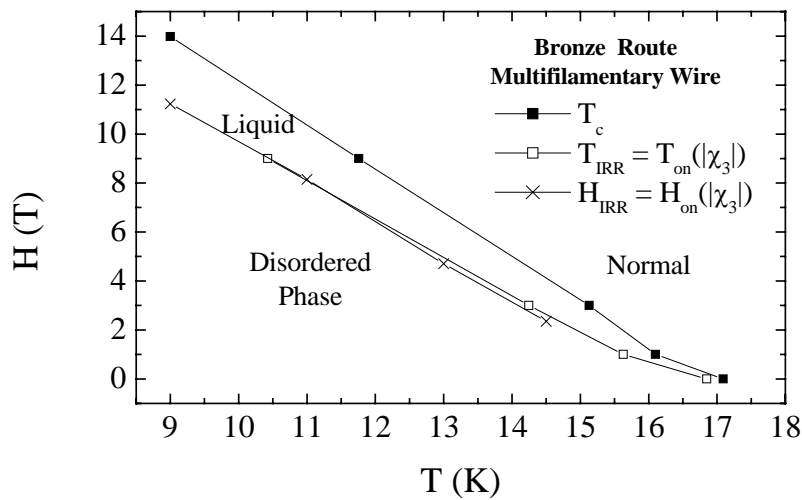

**FIGURE 9**. H-T phase diagram of the multifilamentary bronze route Nb$_3$Sn wire: the Bragg Glass Phase was not detected, but an irreversibility line has been measured.

of the single and the polycrystalline samples.

**CONCLUSIONS**

The magnetic response of three different Nb$_3$Sn samples (a single crystal, a polycrystalline sample and a bronze route multifilamentary wire) has been analyzed by measurements of the 1$^{st}$ and 3$^{rd}$ harmonics of the AC magnetic susceptibility in presence of a static DC field up to 19 Tesla. We observed a common behavior for the single crystal and the polycrystalline samples: a Peak Effect has been identified by 1$^{st}$ harmonics measurements at low fields, and a Bragg Glass phase has been detected up to the maximum available field, by 3$^{rd}$ harmonics. Moreover it has not been possible to distinguish an irreversibility line from the T$_c$-line, in both the samples. A different behavior has been observed in the wire, where an irreversibility line has been detected, in agreement to that reported in the literature [12], but it has not been possible to identify a Bragg Glass Phase. Further differences have also been observed in the non-linear magnetic response, detected by 3$^{rd}$ harmonics measurements; in particular, the wire behavior showed a negative bump in $\chi_3'(T)$ which cannot be described by the critical state models.

**ACKNOWLEDGEMENTS**

We are grateful to N. Toyota, W. Goldacker and V. Abacherli for sending samples. We thank B. Seeber, R. Moudoux and A. Ferreira for helping in the laboratory, and T. Giamarchi for useful discussion.